\documentclass[twocolumn,showpacs,preprintnumbers,amsmath,amssymb]{revtex4}

\usepackage{psfig} 

\usepackage{graphicx} % standard LaTeX graphics tool

\begin{document}

\preprint{APS/123-QED}

\title{Collectives for the Optimal Combination of Imperfect Objects}

\author{Kagan Tumer}
\author{David Wolpert}
\affiliation{NASA Ames Research Center, Moffett Field, CA, 94035, USA\\
{\tt ktumer@mail.arc.nasa.gov \hspace*{.1in} dhw@email.arc.nasa.gov }}

\begin{abstract}

In this letter we summarize some recent theoretical work on the design
of collectives, i.e., of systems containing many agents, each of which
can be viewed as trying to maximize an associated {\bf private
utility}, where there is also a {\bf world utility} rating the
behavior of that overall system that the designer of the collective
wishes to optimize.  We then apply algorithms based on that work on a
recently suggested testbed for such optimization
problems~\cite{chjo02}. This is the problem of finding the combination
of imperfect nano-scale objects that results in the best aggregate
object.  We present experimental results showing that these algorithms
outperform conventional methods by more than an order of magnitude in
this domain.

\end{abstract}

% PACS 


\maketitle              % typesets the title of the contribution

\nocite{joja98,zhan98,huho88,suba98,wada92}

\section{MOTIVATION AND BACKGROUND}
\label{sec:mot}
The optimization problem of finding the subset of a set of imperfect
devices that results in the best aggregate device was recently
introduced by Challet and Johnson~\cite{chjo02}.  It is an abstraction
of what will likely be a major difficulty in the construction of
systems using nano-scale components, arising when a large fraction of
those components may be faulty, so that we cannot use a scheme fixed
ahead of time for interconnecting them.  This abstraction is a
computationally hard optimization problem; brute force approaches
cannot be used for large instances of it.  It is also particularly
well-suited to testing distributed optimization algorithms.

We propose addressing this problem by associating each device with an
adaptive reinforcement-learning (RL) agent~\cite{suba98}) that decides
whether or not its device will be a member of the subset. It makes
this decision based on its estimate of which choice will give a larger
value of its associated {\bf private utility} function which maps the
joint choice of all agents into the reals. For such an approach to
work, we must both ensure that the agents do not work at
cross-purposes, and that each one has a tractable learning problem to
solve. Typically these two desiderata conflict with one another
(utilities that account for whether your action is at cross-purposes
to any other agents' actions are very complex and therefore difficult
to optimize). So we must find the best way to trade them off each
other.

A {\bf collective} is any such system containing utility-maximizing
agents, together with an overall {\bf world utility} function that
rates the possible configurations of the overall system. The
associated design problem is how to configure the collective --- and
in particular how to set the private utility functions --- to best
optimize the world utility.  This design problem is related to work in
many other fields, including multi-agent systems (MAS's),
computational economics, mechanism design, reinforcement learning,
statistical mechanics, computational ecologies, (partially observable)
Markov decision processes and game theory. However none of these
fields is both applicable in large problems, and directly addresses
the {\em general} design problem, rather than a special instance of
it. (See \cite{wotu01a} for a detailed discussion of the relationship
between these fields, involving hundreds of references.) In
particular, since detailed modeling of extremely large real-world
systems is usually impossible, it is crucial to use design algorithms
that tdo not employ such modeling. We need to make our design
leveraging only the simple assumption that the agents' learning
algorithms are individually reasonably behaved. Recently some advances
have been made in doing just that~\cite{tuwo00,wotu01a,wotu99a}. It is
those advances that we propose to apply to the problem of Challet and
Johnson.

\section{The Mathematics of Designing Collectives}
\label{sec:math}

Let $\zeta$ be an arbitrary space whose elements $z$ give the joint
move of all agents in the collective system.  We wish to search for
the $z$ that maximizes the provided world utility $G(z)$. In addition
to $G$ we are concerned with private utility functions \{$g_\eta$\},
one such function for each agent $\eta$ controlling $z_\eta$. We use
the notation $\hat{}\;\eta$ to refer to all agents other than $\eta$.

Our uncertainty concerning the state of the system is reflected in
a probability distribution over $\zeta$. Our ability to control the system
consists of setting the value of some characteristic of the
collective, e.g., setting the private utility functions of the
agents. Indicating that value of the {\bf global coordinate} by $s$,
our analysis revolves around the following {\bf central equation} for
$P(G \mid s)$, which follows from Bayes' theorem:

\begin{eqnarray}
& & P(G \mid s) = \\ \nonumber
& &  \int  d\vec{N}_G P(G \mid \vec{N}_G, s) \int
d\vec{N}_g P(\vec{N}_G \mid \vec{N}_g, s) P(\vec{N}_g \mid s) \; ,
\label{eq:central}
\end{eqnarray}

\noindent
where $\vec{N}_g$ and $\vec{N}_G$ are the {\bf intelligence} vectors
of the agents with respect to $g_\eta$ and $G$,
respectively. Intelligence, defined as the ``standardization'' of
utility functions so that the numeric value they assign to a $z$ only
reflects their ranking of $z$ relative to certain other elements of
$\zeta$, is given by:
\begin{eqnarray}
N_{\eta,U}(z ) \equiv \int d\mu_{z_{\;\hat{}\;\eta}} (z')
	\Theta[U(z) - U(z')] \; ,
\end{eqnarray}

\noindent
where $\Theta$ is the Heaviside function, and where the subscript on
the (normalized) measure $d\mu$ indicates it is restricted to $z'$
sharing the same non-$\eta$ components as $z$. 

Note that $N_{\eta,g_\eta}(z) = 1$ means that agent $\eta$ is fully
rational at $z$, in that its move maximizes the value of its utility,
given the moves of the agents. In other words, a point $z$ where
$N_{\eta,g_\eta}(z) = 1$ for all agents $\eta$ is one that meets the
definition of a game-theory Nash equilibrium.  On the other hand, a
$z$ at which all components of $\vec{N}_G = 1$ is a maximum $G$ along
all coordinates of $z$ (which of course does not mean it is a maximum
along an off-axis direction). So if we can get these two points to be
identical, then if the agents do well enough at maximizing their
private utilities we are assured we will be near an axis-maximizing
point for $G$.

To formalize this, consider our decomposition of $P(G \mid s)$. If we
can choose $s$ so that the third conditional probability in the
integrand is peaked around vectors $\vec{N}_g$ all of whose components
are close to 1, then we have likely induced large (private utility
function) intelligences.  Intuitively, this ensures that the private
utility functions have high ``signal-to-noise''.  If we can also have
the second term be peaked about $\vec{N}_G$ equal to $\vec{N}_g$, then
$\vec{N}_G$ will also be large.  It is in the second term that the
requirement that the private utility functions be ``aligned with $G$''
arises. Note that our desired form for the second term in
Equation~\ref{eq:central} is assured if we have chosen private
utilities such that $\vec{N}_g$ equals $\vec{N}_G$ exactly for all
$z$. Such a system is said to be {\bf factored}.  Finally, if the
first term in the integrand is peaked about high $G$ when $\vec{N}_G$
is large, then our choice of $s$ will likely result in high $G$, as
desired.  In this letter we concentrate on the second and third terms,
and show how to simultaneously set them to have the desired form.

As an example, any ``team game'' in which all private utility
functions equal $G$ is factored~\cite{crba96}.  However team games
often have very poor forms for term 3 in Equation~\ref{eq:central},
forms which get progressively worse as the size of the collective
grows. This is because for such private utility functions each agent
$\eta$ will usually confront a very poor ``signal-to-noise'' ratio in
trying to discern how its actions affect its utility $g_\eta = G$,
since so many other agent's actions also affect $G$ and therefore
dilute $\eta$'s effect on its own private utility function.

We now focus on algorithms based on private utility functions
\{$g_\eta$\} that optimize the signal/noise ratio reflected in the
third term, subject to the requirement that the system be factored.
To understand how these algorithms work, say we are given an arbitrary
function $f(z_\eta)$ over agent $\eta$'s moves, two such moves
${z_\eta}^1$ and ${z_\eta}^2$, a utility $U$, a value $s$ of
the global coordinate, and a move by all agents other than $\eta$,
$z_{\;\hat{}\;\eta}$. Define the associated {\bf learnability} by
\begin{eqnarray}
& & \Lambda_f(U ; z_{\;\hat{}\;\eta}, s, {z_\eta}^1, {z_\eta}^2)
\equiv \\ \nonumber
& & \sqrt{ \frac{ [E(U ; z_{\;\hat{}\;\eta}, {z_\eta}^1) -
 E(U ; z_{\;\hat{}\;\eta}, {z_\eta}^2)]^2}
{\int dz_\eta [f(z_\eta)Var(U ; z_{\;\hat{}\;\eta}, z_\eta)] }} \;
 .
\label{eq:learnability}
\end{eqnarray}
\noindent
The expectation values in the numerator are formed by averaging over
the training set of the learning algorithm used by agent $\eta$,
$n_\eta$. Those two averages are evaluated according to the two
distributions $P(U | n_\eta) P(n_\eta | z_{\;\hat{}\;\eta},
{z_\eta}^1)$ and $P(U | n_\eta) P(n_\eta | z_{\;\hat{}\;\eta},
{z_\eta}^2)$, respectively. (That is the meaning of the semicolon
notation.)  Similarly the variance being averaged in the denominator is
over $n_\eta$ according to the distribution $P(U | n_\eta) P(n_\eta |
z_{\;\hat{}\;\eta}, {z_\eta})$.

The denominator in Equation~\ref{eq:learnability} reflects how
sensitive $U(z)$ is to changing $z_{\;\hat{}\;\eta}$. In
contrast, the numerator reflects how sensitive $U(z)$ is to
changing $z_{\eta}$. So the greater the learnability of a private
utility function $g_\eta$, the more $g_\eta(z)$ depends only on
the move of agent $\eta$, i.e., the better the associated
signal-to-noise ratio for $\eta$. Intuitively then, so long as it does
not come at the expense of decreasing the signal, increasing the
signal-to-noise ratio specified in the learnability will make it
easier for $\eta$ to achieve a large value of its intelligence. This
can be established formally: if appropriately scaled, $g'_\eta$ will
result in better expected intelligence for agent $\eta$ than will
$g_\eta$ whenever $\Lambda_f(g'_\eta; z_{\;\hat{}\;\eta}, s,
{z_\eta}^1, {z_\eta}^2) \; > \; \Lambda_f(g_\eta;
z_{\;\hat{}\;\eta}, s, {z_\eta}^1, {z_\eta}^2)$ for all
pairs of moves ${z_\eta}^1,
{z_\eta}^2$\cite{wolp03a}. 

One can solve for the set of all private utilities that are factored
with respect to a particular world utility. Unfortunately though, in
general a collective cannot both be factored and have infinite
learnability for all of its agents~\cite{wolp03a}.  However consider
{\bf difference} utilities, of the form
\begin{eqnarray}
U(z) = \beta[G(z) - D(z_{\;\hat{}\;\eta})]
\end{eqnarray}
\noindent
Any difference utility is factored~\cite{wolp03a}. In addition, for
all pairs ${z_\eta}^1, {z_\eta}^2$, under benign approximations the
difference utility maximizing $\Lambda_f(U ; z_{\;\hat{}\;\eta}, s,
{z_\eta}^1, {z_\eta}^2)$ by
\begin{eqnarray}
D(z_{\;\hat{}\;\eta}) = 
  E_f(G(z) \mid z_{\;\hat{}\;\eta}, s) \;,
\end{eqnarray}
\noindent
up to an overall additive constant, where the expectation value is
over $z_\eta$. We call the resultant difference utility the {\bf
Aristocrat} utility ($AU$), loosely reflecting the fact that it measures
the difference between a agent's actual action and the average
action.  If each agent $\eta$ uses an appropriately rescaled version
of the associated $AU$ as its private utility function, then we have
ensured good form for both terms 2 and 3 in Equation~\ref{eq:central}.

Using $AU$ in practice is sometimes difficult, due to the need to
evaluate the expectation value. Fortunately there are other utility
functions that, while being easier to evaluate than $AU$, still are both
factored and possess superior learnability to the team game utility,
$g_\eta = G$. One such private utility function is the {\bf Wonderful
Life} Utility (WLU).  The $WLU$ for agent $\eta$ is parameterized by a
pre-fixed {\bf clamping parameter} $CL_\eta$ chosen from among
$\eta$'s possible moves:
\begin{eqnarray}
WLU_\eta \equiv G(z) - G(z_{\;\hat{}\;\eta}, CL_\eta) \; .
\end{eqnarray}
\noindent
WLU is factored no matter what the choice of clamping
parameter. Furthermore, while not matching the high learnability of
$AU$, $WLU$ usually has far better learnability than does a team game, and
therefore (when appropriately scaled) results in better expected
intelligence \cite{tuwo00,wotu99a,wotu01a}.

\section{Combination of Imperfect Objects}
We now explore the use of collective-based techniques for the problem
of combining imperfect objects of Challet and Johnson.  The canonical
version of this problem arises when the objects are all noisy
observational devices producing a single real number by sampling a
Gaussian of fixed width centered on the true value of the number.  The
problem is to choose the subset of a fixed collection of such devices
to have the average (over the members of the subset) distortion as
close to zero as possible.

Formally, the problem is to minimize 
$$
\epsilon \equiv \frac{|
\sum_{j=1}^m n_j a_j |} {\sum_{k=1}^m n_k} \; ,
$$
\noindent
where $n_j \in \{0, 1\}$ is whether device $j$ is or is not selected,
and there are $m$ devices in the collection, having associated
distortions $\{a_j\}$.  We identify $\epsilon$ with the world utility,
$G$ (so that for these experiments, the goal is to minimize $G$, not
maximize it). There are $m$ individual agents, each setting one of the
$n_j$. The goal is to give those agents private utilities so that, as
they learn to maximize their private utilities, the maximizer of $G$
is found.

Because we wished to concentrate on the effects of the utilities
rather than on the RL algorithms that use them, the agents all used
the same (very) simple RL algorithm. (We would expect that even
marginally more sophisticated RL algorithms would give better
performance.)  At each timestep each agent runs its algorithm on a
data set it maintains of action-utility pairs to choose what action to
take. This gives a joint action, which in turn sets the private
utility value for each agent. Combined with what action it took at
that timestep, that utility value for agent $j$ then gets added to the
data set maintained by agent $j$. This is done for all agents and then
the process repeats.  

The agents used their data sets to choose moves by maintaining a
2-dimensional vector whose components at a given timestep are the
agent's estimates of the utility it would receive for taking each of
its two possible move. Each agent $j$ picks its action at a timestep
by sampling a Boltzmann distribution whose over the ``energy
spectrum'' of $j$'s two utility estimates at that time. For
simplicity, given how short our runs were, the temperature in the
Boltzmann distribution did not decay in time. However to reflect the
fact that the environment in which an agent is operating changes with
time (as the other agents change their moves), and therefore the
optimal action changes in time, the two utility estimates are formed
using exponentially aged data: for any time step $t$, the utility
estimate $j$ uses for setting either of the two actions $n_j$ is a
weighted average of all the utility values it has received at previous
times $t'$ that it chose that action, with the weights in the average
given by an exponential of the values $t - t'$.  Finally, to form the
agents' initial data sets, there is an initialization period in which
all actions by all agents are chosen uniformly randomly, with no
learning used. It is after this initialization period ends that the
agents choose their actions according to the associated Boltzmann
distributions.

For all learning algorithms, the first 20 time steps constitute the
data set initialization period (note that all learning algorithms must
``perform'' the same during that period, since none are actually in
use then).  Starting at $t=20$, with each consecutive timestep a fixed
fraction of the agents still choosing their actions randomly switch to
using their learner algorithms instead, while others continue to take
random actions. This gradual introduction of the learning algorithms
is intended to soften the ``discontinuity'' in each agent's
environment when the behavior of the other agents start using their
learning algorithms and therefore change their moves.  For $m=100$,
four agents turned on their learning algorithms at each time step
(i.e., by $t=26$, all agents were using their learners).

\begin{figure} [htb]
%\vspace*{-.2in}
\centerline{\psfig{figure={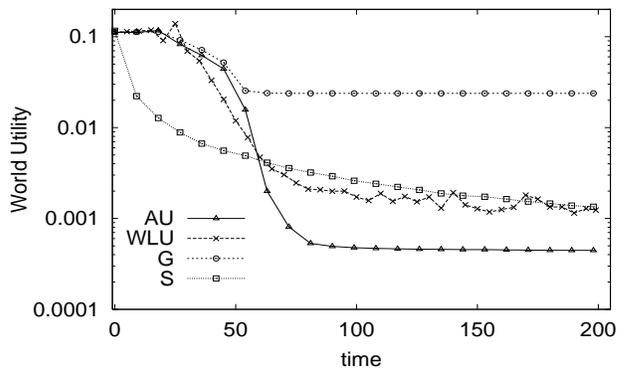},height=2.2in,width=3.4in}}
\caption{Combination of Imperfect Objects Problem, $m=100$.}
\label{fig:n=100}
%\vspace*{-.2in}
\end{figure}

Figure~\ref{fig:n=100} shows the convergence properties of different
algorithms in a system with 100 agents.  The results reported are
based on 20 different $\{a_j\}$ configurations, each performed 50
times (i.e., each point on the figure is the average of $20 \times 50
= 1000$ runs).  $G$, $AU$, and $WLU$ show the performance of agents
using reinforcement learners with those reinforcement signals provided
by $G$ (team game), Aristocrat Utility and Wonderful Life Utility
respectively.  $S$ shows the performance of greedy search where new
$n_j$'s are generated at each step and selected if the solution is
better than the current best solution. Because the runs are only 200
timesteps long, algorithms such as simulated annealing do not
outperform simple search: there is simply no time for an annealing
schedule.

Note that because of the discontinuity in the environment experienced by each
agent as the other agents turn on their learning algorithms, the
performance of the system as a whole as the agents turn on their
learning may degrade initially, before settling down (e.g., WLU).
Qualitatively, systems in which agents use the $G$ utility have a
difficult time learning, while systems in which agents use $AU$
perform best and the search algorithm falls between the performance of
the two.

\begin{figure} [htb]
\centerline{\psfig{figure={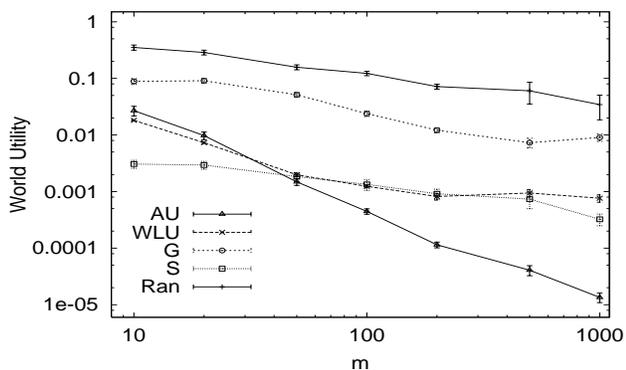},height=2.0in,width=3.4in}}
\caption{Scaling in the Combination of Imperfect Objects Problem.}
\label{fig:scaleW}
%\vspace*{-.2in}
\end{figure}

We also investigated the scaling properties of each algorithm.
Figure~\ref{fig:scaleW} shows these results (the $t=200$ average
performance over 1000 runs) along with the associated error bars.  As
$m$ grows two competing factors come into play.  On the one hand,
there are more degrees of freedom to use to minimize $G$. On the other
hand, the problem becomes more difficult: the search space gets larger
for $S$, and there is more noise in the system for the learning
algorithms. To account for these effects and calibrate the performance
values as $m$ varies, in the figure we also provide the performance of
the algorithm that randomly selects its action (``Ran'').  Note that
the difference between the performances of $S$ and $AU$ increases when
the system size increases, up to a factor of twenty for $m=1000$.

All algorithms but $AU$ have slopes similar to that of
``Ran'', demonstrating that they cannot use the additional
degrees of freedom provided by the larger $m$.  Only $AU$ effectively
uses the new degrees of freedom, providing gains that are
proportionally higher than the other algorithms (i.e., the rate at
which $AU$'s performance improves outpaces what is ``expected'' based
on the random algorithm's performance).

Finally, note that many search algorithms (e.g., gradient ascent,
simulated annealing, genetic algorithms) can be viewed as collectives.
However conventionally such algorithms use very ``dumb'' agents. In
particular, in exploration-exploitation algorithms, the agents
typically make random moves in the exploration step rather than RL to
choose the best move.  Preliminary results indicate that using
RL-based agents to determine the moves in such exploration steps ---
intuitively, ``agentizing'' the individual variables of a search
problem by providing them with adaptive intelligence --- can lead to
significantly better solutions than are achieved with conventional
``dumb variable'' exploration steps in a myriad of optimization
problems~\cite{wotu03a}.

\bibliographystyle{plain}

\end{document}